\documentclass[aps,prd,twocolumn,superscriptaddress,nofootinbib]{revtex4-1}
\pdfoutput=1

\usepackage{float}
\usepackage{amsfonts}
\usepackage{psfrag}
\usepackage{graphicx}
\usepackage{grffile}
\usepackage{cancel}
\usepackage{amsmath}
\usepackage{natbib}
\usepackage{xr}
\usepackage{hyperref}
\usepackage{verbatim}
\usepackage{epstopdf}
\usepackage{subcaption}
\usepackage{caption}
\usepackage{multirow}
\usepackage{float}
\usepackage{commath}
\usepackage{empheq}
\usepackage{braket}
\usepackage{placeins}
\usepackage{ragged2e}

\usepackage{booktabs, dcolumn}
\hypersetup{
    bookmarks=true,         
    unicode=false,          
    pdftoolbar=true,        
    pdfmenubar=true,        
    pdffitwindow=false,     
    pdfstartview={FitH},    
    pdfauthor={Mariska Hoogkamer},     
    colorlinks=true,       
    linkcolor=blue,          
    citecolor=blue        
}
\newcolumntype{d}[1]{D{.}{.}{#1}}

\newcommand\physrep{Physics Reports}

\newcommand\apjl{The Astrophysical Journal Letters}

\newcommand\mnras{Monthly Notices of the Royal Astronomical Society}
\newcommand\aap{Astronomy and Astrophysics}

\RequirePackage[normalem]{ulem} 
\RequirePackage{color}\definecolor{RED}{rgb}{1,0,0}\definecolor{BLUE}{rgb}{0,0,1}

\providecommand{\DIFaddbegin}{} 
\providecommand{\DIFaddend}{} 
\providecommand{\DIFdelbegin}{} 
\providecommand{\DIFdelend}{}

\providecommand{\DIFaddbeginFL}{} 
\providecommand{\DIFaddendFL}{} 
\providecommand{\DIFdelbeginFL}{} 
\providecommand{\DIFdelendFL}{}

\newcommand{\DIFscaledelfig}{0.5}
\RequirePackage{settobox} 
\RequirePackage{letltxmacro} 
\newsavebox{\DIFdelgraphicsbox} 
\newlength{\DIFdelgraphicswidth} 
\newlength{\DIFdelgraphicsheight} 
\LetLtxMacro{\DIFOincludegraphics}{\includegraphics} 
\newcommand{\DIFaddincludegraphics}[2][]{{\color{blue}\fbox{\DIFOincludegraphics[#1]{#2}}}} 
\newcommand{\DIFdelincludegraphics}[2][]{
\sbox{\DIFdelgraphicsbox}{\DIFOincludegraphics[#1]{#2}}
\settoboxwidth{\DIFdelgraphicswidth}{\DIFdelgraphicsbox} 
\settoboxtotalheight{\DIFdelgraphicsheight}{\DIFdelgraphicsbox} 
\scalebox{\DIFscaledelfig}{
\parbox[b]{\DIFdelgraphicswidth}{\usebox{\DIFdelgraphicsbox}\\[-\baselineskip] \rule{\DIFdelgraphicswidth}{0em}}\llap{\resizebox{\DIFdelgraphicswidth}{\DIFdelgraphicsheight}{
\setlength{\unitlength}{\DIFdelgraphicswidth}
\begin{picture}(1,1)
\thicklines\linethickness{2pt} 
{\color[rgb]{1,0,0}\put(0,0){\framebox(1,1){}}}
{\color[rgb]{1,0,0}\put(0,0){\line( 1,1){1}}}
{\color[rgb]{1,0,0}\put(0,1){\line(1,-1){1}}}
\end{picture}
}\hspace*{3pt}}} 
} 
\LetLtxMacro{\DIFOaddbegin}{\DIFaddbegin} 
\LetLtxMacro{\DIFOaddend}{\DIFaddend} 
\LetLtxMacro{\DIFOdelbegin}{\DIFdelbegin} 
\LetLtxMacro{\DIFOdelend}{\DIFdelend} 
\DeclareRobustCommand{\DIFaddbegin}{\DIFOaddbegin \let\includegraphics\DIFaddincludegraphics} 
\DeclareRobustCommand{\DIFaddend}{\DIFOaddend \let\includegraphics\DIFOincludegraphics} 
\DeclareRobustCommand{\DIFdelbegin}{\DIFOdelbegin \let\includegraphics\DIFdelincludegraphics} 
\DeclareRobustCommand{\DIFdelend}{\DIFOaddend \let\includegraphics\DIFOincludegraphics} 
\LetLtxMacro{\DIFOaddbeginFL}{\DIFaddbeginFL} 
\LetLtxMacro{\DIFOaddendFL}{\DIFaddendFL} 
\LetLtxMacro{\DIFOdelbeginFL}{\DIFdelbeginFL} 
\LetLtxMacro{\DIFOdelendFL}{\DIFdelendFL} 
\DeclareRobustCommand{\DIFaddbeginFL}{\DIFOaddbeginFL \let\includegraphics\DIFaddincludegraphics} 
\DeclareRobustCommand{\DIFaddendFL}{\DIFOaddendFL \let\includegraphics\DIFOincludegraphics} 
\DeclareRobustCommand{\DIFdelbeginFL}{\DIFOdelbeginFL \let\includegraphics\DIFdelincludegraphics} 
\DeclareRobustCommand{\DIFdelendFL}{\DIFOaddendFL \let\includegraphics\DIFOincludegraphics} 

\newcommand{\Msun}{$M_\odot$}

\begin{document}

\title{Cross-Comparison of Sampling Algorithms\\ for Pulse Profile Modeling of PSR J0740+6620}
\author{Mariska Hoogkamer}
\affiliation{Anton Pannekoek Institute for Astronomy, University of Amsterdam, Science Park 904, 1098XH Amsterdam, the Netherlands}

\author{Yves Kini}
\affiliation{Anton Pannekoek Institute for Astronomy, University of Amsterdam, Science Park 904, 1098XH Amsterdam, the Netherlands}

\author{Tuomo Salmi}
\affiliation{Department of Physics, University of Helsinki, P.O. Box 64, FI-00014, Finland}

\author{Anna L. Watts}
\affiliation{Anton Pannekoek Institute for Astronomy, University of Amsterdam, Science Park 904, 1098XH Amsterdam, the Netherlands}

\author{Johannes Buchner}
\affiliation{Max Planck Institute for Extraterrestrial Physics, Giessenbachstrasse 1, 85748 Garching, Germany}

\date{\today}

\begin{abstract}
In the last few years, NICER data has enabled mass and radius inferences for various pulsars, and thus shed light on the equation of state for dense nuclear matter. This is achieved through a technique called pulse profile modeling. The importance of the results necessitates careful validation and testing of the robustness of the inference procedure. In this paper, we investigate the effect of sampler choice for X-PSI (X-ray Pulse Simulation and Inference), an open-source package for pulse profile modeling and Bayesian statistical inference that has been used extensively for analysis of NICER data.
We focus on the specific case of the high-mass pulsar PSR J0740+6620. Using synthetic data that mimics the most recently analyzed NICER and XMM-Newton data sets of PSR J0740+6620, we evaluate the parameter recovery performance, convergence, and computational cost for \verb|MultiNest|'s multimodal nested sampling algorithm and \verb|UltraNest|'s slice nested sampling algorithm. We find that both samplers perform reliably, producing accurate and unbiased parameter estimation results when analyzing simulated data. We also investigate the consequences for inference using the real data for PSR J0740+6620, finding that both samplers produce consistent credible intervals.  
\end{abstract}

\maketitle

\section{Introduction}\label{sec:introduction}
Neutron stars are among the densest objects in the Universe, making them excellent for studying the behavior of dense nuclear matter and constraining the equation of state (EoS) \cite[see e.g.,][]{Lattimer2016-EoS, Baym2018-EoS, Tolos2020}. One effective approach to probing neutron star properties, and thereby constraining the EoS, is through modeling the X-ray emission from their surfaces. The Neutron Star Interior Composition Explorer (NICER; \cite{Gendreau2016-NICER}), located onboard the International Space Station, has been pivotal in this effort by detecting soft thermal X-rays from rotation-powered millisecond pulsars. These pulsars are a distinct category of neutron stars, showing rotationally modulated and hence pulsed X-ray emission that is thought to originate from heat deposited at the magnetic poles by return currents \cite[see e.g.,][]{Ruderman1975-pulsars, Arons1981, Harding2001}. By analyzing NICER observations using pulse profile modeling \cite[e.g.,][and references therein]{Watts2019-ppm,Bogdanov2019-part2, Bogdanov2021-part3} --- a technique that incorporates relativistic effects caused by the neutron star's rapid spin and strong gravitational field --- it is possible to derive precise measurements of neutron star masses and radii. These measurements, in turn, provide stringent constraints on the EoS governing cold, ultra-dense matter \cite[see e.g.][]{Rutherford24,Koehn24,Huang25,Golomb25,LiJJ24}. 

In the last few years, NICER data has enabled mass and radius inferences for four pulsars: PSR J0030+0451, \cite{Riley2019-J0030-xpsi, Miller2019-J0030, Salmi2023-atmosphere, Vinciguerra2024-J0030}, PSR J0437-4715 \cite{Choudhury2024-J0437}, PSR J0740+6620 \cite{Miller2021-J0740, Riley2021-J0740, Salmi2022-J0740, Salmi2023-atmosphere, Salmi2024-J0740, Dittmann2024-J0740}, and PSR J1231-1411 {\cite{Salmi2024-J1231}. In this work, we focus on the high-mass pulsar PSR J0740+6620 using the most recently analyzed NICER data set (from 2018 September 21 to 2022 April 21), and XMM-Newton data (this is included to provide indirect constraints on the background), as studied by \cite{Salmi2024-J0740}. The inferred equatorial radius and gravitational mass found were $12.49^{+1.28}_{-0.88}$ km and $2.073^{+0.069}_{-0.069}$ \Msun \cite{Salmi2024-J0740}, with the latter being largely determined by the prior from radio pulsar timing \cite{Fonseca2021-J0740}. 

This inferred mass and radius estimate (along with several of the other results cited previously) was obtained using the X-ray Pulse Simulation and Inference (X-PSI) code\footnote{\url{https://github.com/xpsi-group/xpsi}}, which is a software package for pulse profile modeling and Bayesian statistical inference \cite{xpsi}. To date, it has been used in combination with the sampler \verb|MultiNest|\cite{Feroz2008-MultimodalNestedSampling, Feroz2009-Multinest, Feroz2019-Multinest} and its Python bindings \verb|PyMultiNest|\cite{Buchner2014-Pymultinest}. 

The inferred equatorial radius interval of $12.49^{+1.28}_{-0.88}$ km found by \cite{Salmi2024-J0740} is slightly different to the inferred radius interval of $12.76^{+1.49}_{-1.02}$ km found by \cite{Dittmann2024-J0740} using a different pulse profile modeling procedure (and when setting the upper limit on the radius prior to 16 km for consistency with \cite{Salmi2024-J0740}). The difference between the results could be related to different sampling procedures (see Section 4.3 of \cite{Salmi2024-J0740} for a comparison of the results with \cite{Dittmann2024-J0740}). 

Discrepancies between results obtained with different samplers are not uncommon. For example, the analysis of gravitational waves from merging compact objects yields different results in terms of robustness, efficiency of producing posterior samples, and accuracy of estimating the evidence \cite{Ashton2019-Bilby, Ashton2021-BilbyMCMC}. This highlights the importance of cross-sampler comparisons.   

Not only is the choice of sampling algorithm important, but also the choice of sampler settings \cite{Vinciguerra2023-J0030-sim}. Sampler settings regulate, among other things, how exhaustively the parameter space is explored, potentially causing biases. Proving convergence is essential to mitigate systematics and ensure the robustness of the results. Ensuring that the parameter space has been thoroughly explored can, for example, be done by carrying out runs with more and more stringent sampling requirements, as well as repeated inferences, assessing the variability due to the randomness of the process involved \cite{Vinciguerra2023-J0030-sim}. 

In this paper, we aim to explore the effect of different sampler choices within X-PSI to test the robustness of previous inference results. We focus on the specific case of PSR J0740+6620, building upon the analysis of \cite{Salmi2024-J0740}. We test two different sampling algorithms, starting with the multimodal nested sampling algorithm implemented in \verb|MultiNest|, as used in our earlier works \cite[see e.g.,][]{Riley2019-J0030-xpsi, Salmi2022-J0740, Salmi2024-J0740, Choudhury2024-J0437}. We compare this to the slice sampling algorithm \cite{Neal2003-SliceSampling, Buchner2022-ComparisonStepSamplers} implemented in \verb|UltraNest| \cite{Buchner2021-UltraNest}. \verb|UltraNest| is a sampler that is generally more robust and known to be less prone to biases (see Section \ref{subsec:nested_sampling}), albeit generally more computationally expensive than \verb|MultiNest| \cite{Buchner2021-UltraNest}. We test the performance of the sampling algorithms by performing parameter recovery tests with simulated data. We track and compare the computational costs, and then investigate the effect of our sampler choice on inference using the real PSR J0740+6620 data set.  

The remainder of the paper is structured as follows: in Section \ref{sec:Methodology} we summarize the methodology and describe the nested sampling algorithm including the two variants used in this paper. Next, in Section \ref{sec:data} we describe the simulated X-ray event data and the real PSR J0740+6620 data. In Section \ref{sec:Results} we state our results and lastly, we discuss our findings and future work in Section \ref{sec:Discussion}. 

\section{Methodology}\label{sec:Methodology}
Our approach to modeling the X-ray data and our parameter estimation is similar in most aspects to the procedures outlined in \cite{Riley2021-J0740, Salmi2022-J0740, Salmi2024-J0740}. We briefly summarize the most important aspects in Section \ref{subsec:xpsi} and how it differs from our previous work. We then describe the nested sampling algorithm and the two variants used in this paper in Section \ref{subsec:nested_sampling}. Complete information of each run, including the exact X-PSI version, data products, posterior sample files, and analysis files can be found in the Zenodo repository of \cite{Hoogkamer2025-Zenodo}. 

\subsection{Pulse Profile Modeling using X-PSI} \label{subsec:xpsi}
We use the open-source X-PSI package, with versions ranging from \verb|v2.2.0| to \verb|v2.2.7| \cite{xpsi}. As done previously \cite[see e.g.,][]{Riley2021-J0740, Salmi2022-J0740, Salmi2024-J0740}, we use the `Oblate Schwarzschild + Doppler' approximation to model the energy-resolved X-ray pulses from the neutron star \cite{Miller1998, Nath2002, Poutanen2003, Morsink2007, Lo2013, AlGendy2014, Bogdanov2019-part2, Watts2019-ppm}. This approximation takes the oblate shape of the neutron star into account and the relativistic effects resulting from the rapid spin of the rotation-powered millisecond pulsars while treating the exterior spacetime as Schwarzschild. See \cite{Riley2019-J0030-xpsi, xpsi} for a more detailed description of X-PSI.  

For PSR J0740+6620, we use the priors for mass, inclination, and distance from \cite{Fonseca2021-J0740}. Additionally, we use the \verb|TBabs| model (\cite{Wilms2000-xrayISM}, updated in 2016) to account for the absorption of X-rays in the interstellar medium. We use the fully ionized hydrogen atmosphere model \verb|NSX| \cite{Ho2001-atmosphere, Ho2009-atmosphere}, an assumption which does not seem to have a significant effect on the inferred radius of PSR J0740+6620 \cite{Salmi2023-atmosphere}. Furthermore, the hot emitting regions of the neutron star are modeled with two circular uniform temperature regions using the \verb|ST-U| (\textit{Single-Temperature-Unshared}) model (see \cite{Vinciguerra2023-J0030-sim, Riley2019-J0030-xpsi} for more details on the X-PSI model naming convention and parameters). To compute the likelihood we adopt either high or low-resolution settings \footnote{\label{footnote:res_settings} \raggedright Low-resolution settings: \texttt{num\_leaves}=32, \texttt{num\_energies}=32, \texttt{sqrt\_num\_cells}=18, and \texttt{max\_sqrt\_num\_cells}=32. High-resolution settings: \texttt{num\_leaves}=64, \texttt{num\_energies}=128, \texttt{sqrt\_num\_cells}=32, and \texttt{max\_sqrt\_num\_cells}=64. } for our inference runs (for details see see Section 2.3.1 of \cite{Vinciguerra2023-J0030-sim}). These settings control the discretization of the computational domain for computation of signals (pulse profiles) incident on the telescope. In this study, we mostly opt for low-resolution settings, because they significantly reduce computational cost while having little to no noticeable impact on the resulting posteriors.  

\subsection{Nested Sampling} \label{subsec:nested_sampling}
X-PSI is used in combination with a sampling algorithm to explore the parameter space. More specifically, it commonly uses the nested sampling algorithm originally designed by Skilling in 2004 \cite{Skilling2004-NestedSampling}. This is a Monte Carlo method that generates parameter posterior samples and is used for Bayesian model comparison by calculating the evidence \cite{Skilling2004-NestedSampling, Skilling2006-NestedSampling2}.

Nested sampling is an iterative integration procedure that shrinks the prior volume towards higher likelihoods. During initialization, $N$ live points are randomly sampled from the prior space. The likelihood $L$ at each point is evaluated. The live point with the lowest likelihood is removed, shrinking the prior volume by a factor of approximately $\delta V=1/N$. A new live point is then drawn uniformly from the prior, with the requirement that its likelihood exceeds that of the live point it replaces, i.e. $L > L_{\mathrm{min}}$, where $L_{\mathrm{min}}$ is the likelihood threshold. This process is known as sampling under a constrained prior, or constrained sampling for short \cite{Buchner2014-test}. This process is repeated $i$ times after which the remaining volume is exponentially small, approximately $V_i=(1-1/N)^i$, with a high likelihood threshold selecting live points close the the best-fit parameters. The algorithm terminates and is said to be converged when further iterations would not significantly alter the result \cite{Buchner2023-NestedMethods}. 

\begin{figure*}[ht!]
\centering
\includegraphics[
width=\textwidth]
{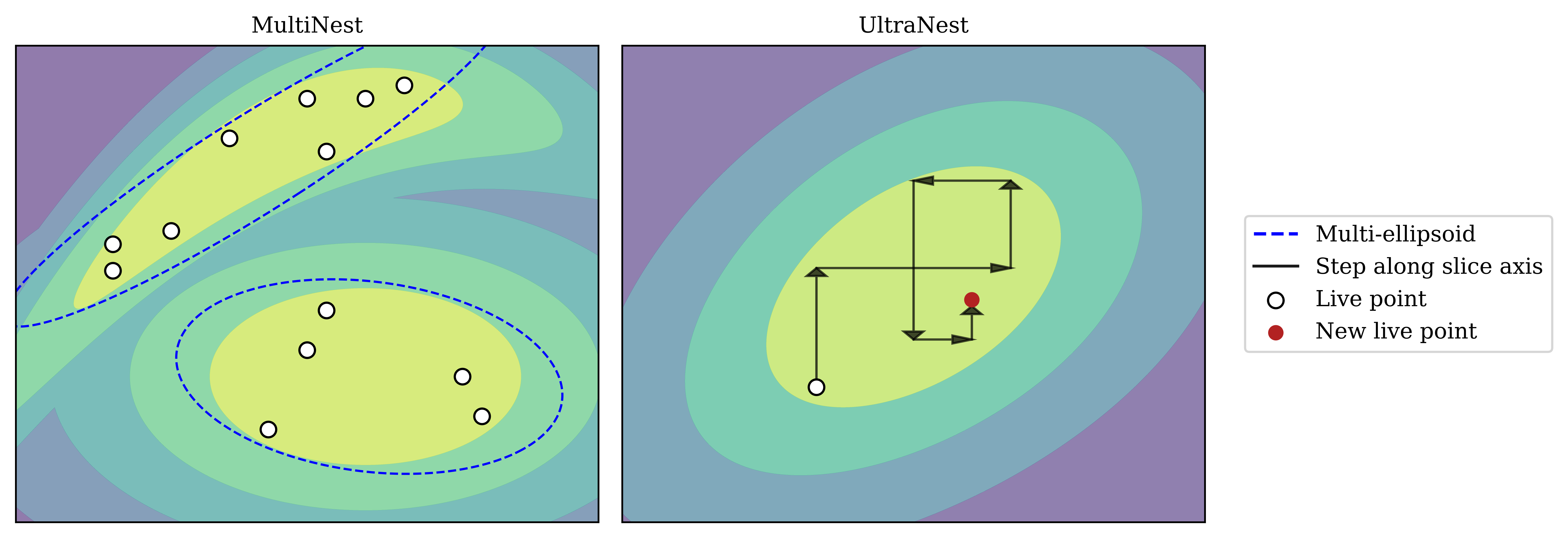}  
\caption{\small \justifying Illustration of two different methods for identifying a new live point (white circles) within an arbitrary parameter space. \textit{Left panel:} multi-ellipsoidal nested sampling, as implemented in \texttt{MultiNest}, approximates the unknown likelihood surface (colored contours) by constructing multi-ellipsoidal regions (blue dashed lines) around the current set of live points (white circles). Rejection sampling based on these contours can become inefficient if the contours are too large, and problematic for nested sampling integration if a region is missed (e.g., the top-right yellow tail). For clarity, the enlargement of the ellipsoids is intentionally chosen to be too small here. \textit{Right panel:} slice sampling, as implemented in \texttt{UltraNest}, performs a Metropolis-like random walk (black arrows) starting from an existing live point (white circle) along ``slice" axes. Steps that fall outside the likelihood contour are rejected. After a sufficient number of steps and with well-tuned proposals, a new live point (red circle) is identified that is independent of the initial point. 
}
\label{fig:nested_sampling}
\end{figure*}

\subsubsection{MultiNest}
One of the samplers that is often used in combination with X-PSI is \verb|MultiNest| and its Python bindings \verb|PyMultiNest| \cite{Buchner2016-PyMultiNest} (for example \cite{Riley2019-J0030-xpsi, Salmi2024-J0740, Choudhury2024-J0437}). \verb|MultiNest| is a multimodal nested sampling algorithm \cite{Feroz2008-MultimodalNestedSampling} that calculates the evidence, with an associated error estimate, and produces posterior samples from distributions that may contain multiple modes and pronounced (curving) degeneracies in high dimensions \cite{Feroz2009-Multinest}. The latter is what sets the multimodal nested sampling algorithm apart from the ``classic" nested sampling algorithm which is only efficient for unimodal distributions without pronounced degeneracies \cite{Mukherjee2006}. 

\verb|MultiNest| uses region sampling (see Section 5.2 of \cite{Buchner2023-NestedMethods} for more details) in order to draw new live points from the prior but above a certain likelihood threshold. In region sampling, it is guessed where the permitted region lies, and a new live point is drawn directly from the prior. The guess is augmented by live points tracing out the likelihood contour. In \verb|MultiNest|, this is done with a clustering algorithm that encapsulates the live points in a number of hyperellipses and draws only from inside these regions \cite{Buchner2014-test} as illustrated in the left panel of Figure \ref{fig:nested_sampling}.   

The \verb|MultiNest| algorithm is mainly controlled by two main parameters:
\begin{enumerate}
    \item Number of live points: this determines the number of samples that are initially drawn from the prior volume. Later, these are replaced according to the procedure briefly described above (for more detail see \cite{Feroz2009-Multinest}). 
    \item Sampling efficiency (or equivalently the inverse of the expansion factor 1/$e$): this parameter sets the enlargement factor of the prior volume during sampling \cite{Feroz2009-Multinest}. This enlargement factor is used to widen the prior volume defined by the clusters (ellipsoids) and to try to ensure that the entire isolikelihood contours are enclosed. In X-PSI the sampling efficiency is scaled by the fraction of the unit hypercube sampling space effectively allowed by the prior conditions and rejection rules (see Appendix B of \cite{Riley2019-J0030-xpsi} and Appendix B of \cite{Salmi2024-J0740} for more details on its implementation in X-PSI). 
\end{enumerate}

Accuracy and precision of evidence estimates and posterior distributions increase with low sampling efficiency and high number of live points. Analysis of a set of simple, analytically tractable test problems by \cite{Dittmann2024-Multinest-notes} has suggested that if one finds consistent evidence estimates and posterior distributions between two or more \verb|MultiNest| analyses that have varied the number of live points and sampling efficiencies by factors of $\sim$three or more, the results of the analyses are likely converged and can be trusted. However, the computational cost of the analysis also increases with lower sampling efficiency and a higher number of live points. This makes it sometimes computationally not feasible to formally prove convergence, as is the case for \cite{Riley2021-J0740, Salmi2024-J0740}.

All in all, \verb|MultiNest| is a widely used sampler, however, there are some drawbacks. As mentioned before, \verb|MultiNest| uses region sampling to draw new live points from a likelihood constricted space. Two main problems can occur when the sampling region is made:
\begin{enumerate}
    \item The sampling region contains space that falls below the likelihood threshold $L_{\mathrm{min}}$. Consequently, the sampled points are useless and have to be rejected. This is especially problematic in higher dimensional problems where this space is larger ($D \gtrsim 20$) \cite{Feroz2008-MultimodalNestedSampling}.
    \item The sampling region misses space that falls within the likelihood threshold. To counteract this problem, the sampling region is expanded by the factor $e$. However, this does not always suffice. When the prior space is underestimated, it can lead to biased likelihoods, to either higher or lower values \cite{Buchner2014-test, Buchner2023-NestedMethods, Dittmann2024-Multinest-notes}.  An example of this is shown in the left panel of Figure \ref{fig:nested_sampling}. 
\end{enumerate}

For more details on \verb|MultiNest| see \cite{Feroz2008-MultimodalNestedSampling, Feroz2009-Multinest}, and see Section 2.4 of \cite{Vinciguerra2023-J0030-sim}. 

\subsubsection{UltraNest} 
\verb|UltraNest| is another Bayesian inference package designed for parameter estimation and model comparison. It is designed to prioritize correctness and robustness \footnote{In this context, correctness refers to the sampler accurately computing the posterior and evidence, while robustness refers to the sampler performing reliably across diverse and complex problem spaces.}, and then speed \cite{Buchner2021-UltraNest}. \verb|UltraNest| has many different sampling algorithms that are implemented, but in this paper we focus on the slice sampling algorithm \cite{Neal2003-SliceSampling, Buchner2022-ComparisonStepSamplers}. 

The slice sampler is a type of Monte Carlo random walk, which is especially useful for high dimensional problems ($D>20$)\cite{Buchner2021-UltraNest}. Slice sampling works by uniformly drawing a new point in the vertical direction alternating with uniformly sampling from the horizontal ``slice" defined by the current vertical position, which is illustrated in the right panel of Figure \ref{fig:nested_sampling}. After a number of Metropolis steps, for which points with lower likelihood are off limits, a new prior sample is obtained. This method is only effective if enough steps are made to ensure that all relevant parameter space is explored \cite{Buchner2024-RelativeJump}.  

\verb|UltraNest|'s slice sampling algorithm is mainly controlled by the following parameters:
\begin{enumerate}
    \item Number of live points: the number of samples that are initially drawn from the prior volume. 
    \item Number of steps ($N_\mathrm{steps}$): this determines how often a geometric random walk is started from a randomly chosen live point before replacing it as starting point for the next iteration (as long as it exceeds the current likelihood threshold). It should be chosen to be large enough so that the final point is sufficiently independent from the starting point \cite{Salomone2023-NestedSampling, Buchner2024-RelativeJump}. 
    \item Proposal function: this determines the direction in which the next sample is chosen. We use \verb|generate_mixture_random_direction| which is the best method according to \cite{Buchner2022-ComparisonStepSamplers}. It is a proposal that samples randomly and uniformly from two other proposals. The first proposal samples from a vector using the difference between two randomly selected live points. The second proposal samples from a vector along one region principal axis, chosen at random. 
\end{enumerate}

\subsubsection{MultiNest versus UltraNest}
The main difference between \verb|MultiNest| and \verb|UltraNest| is that they rely on different sampling algorithms. The multi-ellipsoidal nested sampling algorithm of \verb|MultiNest| is a type of region sampler, whereas the slice sampling algorithm of \verb|UltraNest| is a type of step sampler. For a visual comparison of these sampling methods, see Figure \ref{fig:nested_sampling} or refer to Figure 8 of \cite{Buchner2023-NestedMethods}. The advantage of a step sampler is that it escapes the curse of dimensionality as their cost only shows polynomial $\mathcal{O}(D^b)$ scaling with dimensionality, where $b$ is the anticipated dimensional scaling. Nevertheless, region samplers are often more efficient in low dimensions. Therefore, step samplers are more often used with high dimensional problems ($D\gtrsim20$) \cite{Ashton2022-NestedSampling}. 

Another major difference between the samplers is that \verb|UltraNest| uses reactive nested sampling, thus adapting the number of live points during the inference run (see \cite{Buchner2021-UltraNest} for more details). The advantage of using a dynamic number of live points is that it reduces the uncertainties in the inference \cite{Ashton2022-NestedSampling}. \verb|MultiNest| uses a fixed number of live points throughout the run, which needs careful study to ensure that this number is sufficient for the problem at hand (see e.g. \cite{Vinciguerra2023-J0030-sim}).

Proving convergence for \verb|MultiNest| can be done by checking if the posteriors and evidence are stable when increasing the number of live points and decreasing the sampling efficiency \cite{Higson2018-nestcheck}. Similarly, for \verb|UltraNest|'s slice sampler it can be done with increasing the initial number of live points, or increasing the number of steps. 

In addition to assessing posteriors and evidence for signs of convergence by increasing sampler settings over multiple runs, \verb|UltraNest| contains self-diagnosing algorithms to show whether a (single) run has converged. The first one is based on determining the relative jump distance, $\mathrm{RJD}$, which is the Mahalanobis distance between the start and end point of a random walk divided by the typical neighbor distance between live points. If $\mathrm{RJD}>1$ for the majority of the samples, the parameter space is sufficiently explored, and the results can be trusted\cite{Buchner2024-RelativeJump}. 

Another way \verb|UltraNest| quantifies the convergence of a run is through the Mann-Whitney-Wilcoxon U-test (see Section 4.5.2 of \cite{Buchner2023-NestedMethods} or \cite{Fowlie2020-Utest} for details). This test assesses whether the insertion order of new live points is uniformly distributed. If the samples are correlated or biased instead of independent, the insertion orders will deviate from uniformity, causing the U-test to fail.  The advantage of these self-diagnostic tools is that the run can be evaluated independently, without requiring comparison to other runs, thus significantly reducing computational costs.

\verb|MultiNest| implements and defaults to multi-ellipsoidal rejection. This is a heuristic technique that is known to give biased results for several test problems \cite{Dittmann2024-Multinest-notes, Buchner2014-test, Beaujean2013}. For X-PSI, \cite{Vinciguerra2023-J0030-sim} found an example where imposing a tight prior constraint on inclination also led to biased results. This problem does not really go away with decreasing sampling efficiency (increasing the expansion factor $e$). \verb|UltraNest|'s slice sampling algorithm implements safer uncertainty estimation than \verb|MultiNest|. It incorporates the scatter in both volume estimates and likelihood estimates, whereas \verb|MultiNest| only supports a static volume uncertainty estimate. The nested integrator of \verb|UltraNest|, when assigning weights to the sampled points, uses a bootstrapping scheme that simulates other runs with fewer live points in order to get more robust and realistic uncertainties.

All in all, \verb|UltraNest| is a sampler that is generally more robust and known to be less prone to biases, albeit generally more computationally expensive than \verb|MultiNest| \cite{Buchner2021-UltraNest}.

\section{X-ray Event Data}\label{sec:data}
In this section, we outline the X-ray event data used in this study. In Section \ref{subsec:real_data}, we describe the most recently analyzed NICER data set, spanning from 2018 September 21 to 2022 April 21, along with the XMM-Newton observations of PSR J0740+6620, as detailed in \cite{Salmi2024-J0740}. Section \ref{subsec:synthetic_data} focuses on the synthetic data, designed to replicate these observations, which are used for the parameter recovery tests. 

\subsection{Real Data of PSR J0740+6620}\label{subsec:real_data}
We use the NICER data of PSR J0740+6620 collected between September 21, 2018, and April 21, 2022 (observation IDs 1031020101 through 5031020445). The NICER X-ray event data is processed in the exact same manner as reported in \cite{Salmi2024-J0740}. After filtering, the final dataset includes 2.73 Ms of on-source exposure time, representing an increase of more than 1 Ms compared to the ``old" NICER dataset (2018 September 21 $-$ 2020 April 17) used in \cite{Riley2021-J0740, Miller2021-J0740}.

For the XMM-Newton data, we utilize the same phase-averaged spectral data and blank-sky observations used for background constraints with the three EPIC instruments (pn, MOS1, MOS2) as reported in \cite{Riley2021-J0740, Wolff2021, Salmi2022-J0740, Salmi2024-J0740}.

\subsection{Synthetic Data mimicking PSR J0740+6620}\label{subsec:synthetic_data}
To evaluate the performance of the sampling algorithms, we conduct several inference runs using synthetic data. We generate ten synthetic datasets for NICER and XMM-Newton using ten distinct parameter vectors, which are detailed in the Appendix. These parameter vectors are randomly sampled from the prior distributions employed in the joint NICER and XMM-Newton analysis of PSR J0740+6620 (see Table I, column 3 of \cite{Salmi2024-J0740} for details about the priors), which is also what is used for the inference runs. Note that in our case, the effective-area scaling factors, $\alpha_{\rm{NICER}}$ and $\alpha_{\rm{XMM}}$, are fixed to 1.0 since they are not expected to influence the results.   

The exposure times are kept identical to those of the actual observation used in \cite{Salmi2024-J0740} (see Section \ref{subsec:real_data}). The input background for each instrument is set to the background that maximizes the instrument-specific likelihood for the real data (see \cite{Salmi2024-J0740} for details). Poisson fluctuations are added to the model counts using a different random seed for the generator. 

\section{Results} \label{sec:Results}
The results, aimed at showing the effect of our sampler choice within X-PSI, are presented in two parts. In Section \ref{subsec:param_recovery}, we show the results of parameter recovery tests on simulated data, which are used to evaluate the performance of two of the sampling algorithms. In Section \ref{subsec:inference}, we present the inference results of both samplers on the real data of PSR J0740+6620. 

\begin{figure*}[ht]
\centering
\includegraphics[width=0.9\textwidth]{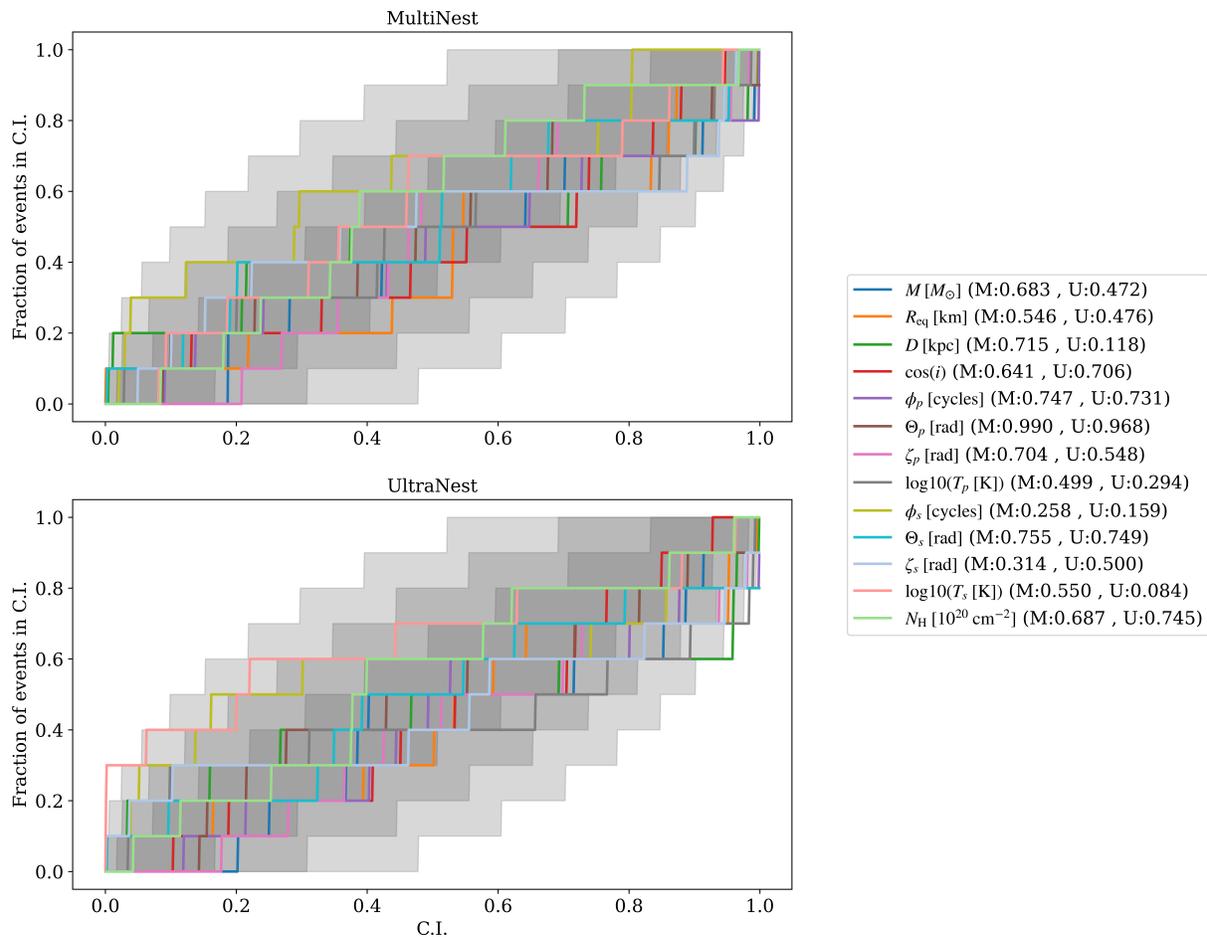}
\caption{\small \justifying{ 
P-P plot with ten synthetic datasets (as described in Section \ref{subsec:synthetic_data}). \textit{Upper panel:} showing the parameter recovery performance of \texttt{MultiNest} using $4\times10^3$ live points and a sampling efficiency of 0.1. The gray regions cover the cumulative 1-, 2-, and 3-$\sigma$ credible intervals in order of decreasing opacity. Each colored line tracks the cumulative fraction of events within this credible interval for a different parameter, including the individual parameter p-values displayed in parentheses in the plot legend, with M for \texttt{MultiNest} and U for \texttt{UltraNest}. The combined p-value for \texttt{MultiNest} is 0.976. \textit{Lower panel:} same as the upper panel, but results are shown for \texttt{UltraNest} using a minimum of 400 live points and 240 steps. The combined p-value for \texttt{UltraNest} is 0.607.}}
\label{fig:pp_plots}
\end{figure*}

\subsection{Parameter Recovery Tests} \label{subsec:param_recovery}
Parameter recovery tests are commonly performed to assess the performance of sampling algorithms \cite[see e.g.,][]{Cook2012-ValidationPP, Talts2018-ValidationPP, Romero_Shaw2020-Bilby}. These tests verify whether the injected parameter values are recovered within statistically expected credible intervals; for instance, 10\% of the values should lie within the 0.1 credible interval, 60\% within the 0.6 credible interval, and so on \cite{Cook2012-ValidationPP, Talts2018-ValidationPP}.

To evaluate the robustness of X-PSI parameter recovery, we simulate 10 pulse profiles mimicking PSR J0740+6620 (see Section \ref{subsec:synthetic_data}). Parameter estimation is performed using both \verb|MultiNest| and the slice sampling algorithm of \verb|UltraNest|. We then perform a Kolmogorov–Smirnov test \cite{Kolmogorov1933}, representing the probability that the fraction of events in a given confidence interval is drawn from a uniform distribution under the assumption of Poisson likelihoods. The combined p-value indicates the probability that all parameters collectively follow a uniform distribution. We deem the test passed if the combined p-value is greater than 0.01.

To conserve computational resources, we use low-resolution settings (see Footnote \ref{footnote:res_settings}) for the X-PSI likelihood computation and sampler settings that are typically applied in exploratory runs. For \verb|MultiNest|, we use $4\times10^3$ live points and a sampling efficiency of 0.1; similar settings were employed for PSR J0740+6620 in \cite{Salmi2023-atmosphere, Salmi2024-J0740}. For \verb|UltraNest|'s slice sampler, we use a minimum of 400 live points and 240 steps, based on prior test runs.

The results of the tests, in the form of a PP-plot, are shown in Figure \ref{fig:pp_plots}. In this plot, the fraction of events for which the true parameter lies within a particular confidence level is plotted against that confidence interval. The plot also includes p-values for each parameter. 

For \verb|MultiNest|, the combined p-value is 0.976, with a minimum of 0.258 for $\phi_s$. For \verb|UltraNest|'s slice sampler, the combined p-value is 0.607, with a minimum of 0.084 for $\log_{10}(T_s)$. These results show that both samplers perform well in recovering parameters, having a p-value larger than 0.01. Additionally, we expect the colored lines representing individual parameters to deviate from the gray regions—representing cumulative 1-, 2-, and 3-$\sigma$ credible intervals—approximately 0.3\% of the time, which is in line with what we see.

In terms of computational cost, a \verb|MultiNest| run with the specified settings required an average of 4.8k core hours, compared to 45k core hours for \verb|UltraNest|.

\begin{figure}[ht!]
\centering
\includegraphics[width=\columnwidth]
{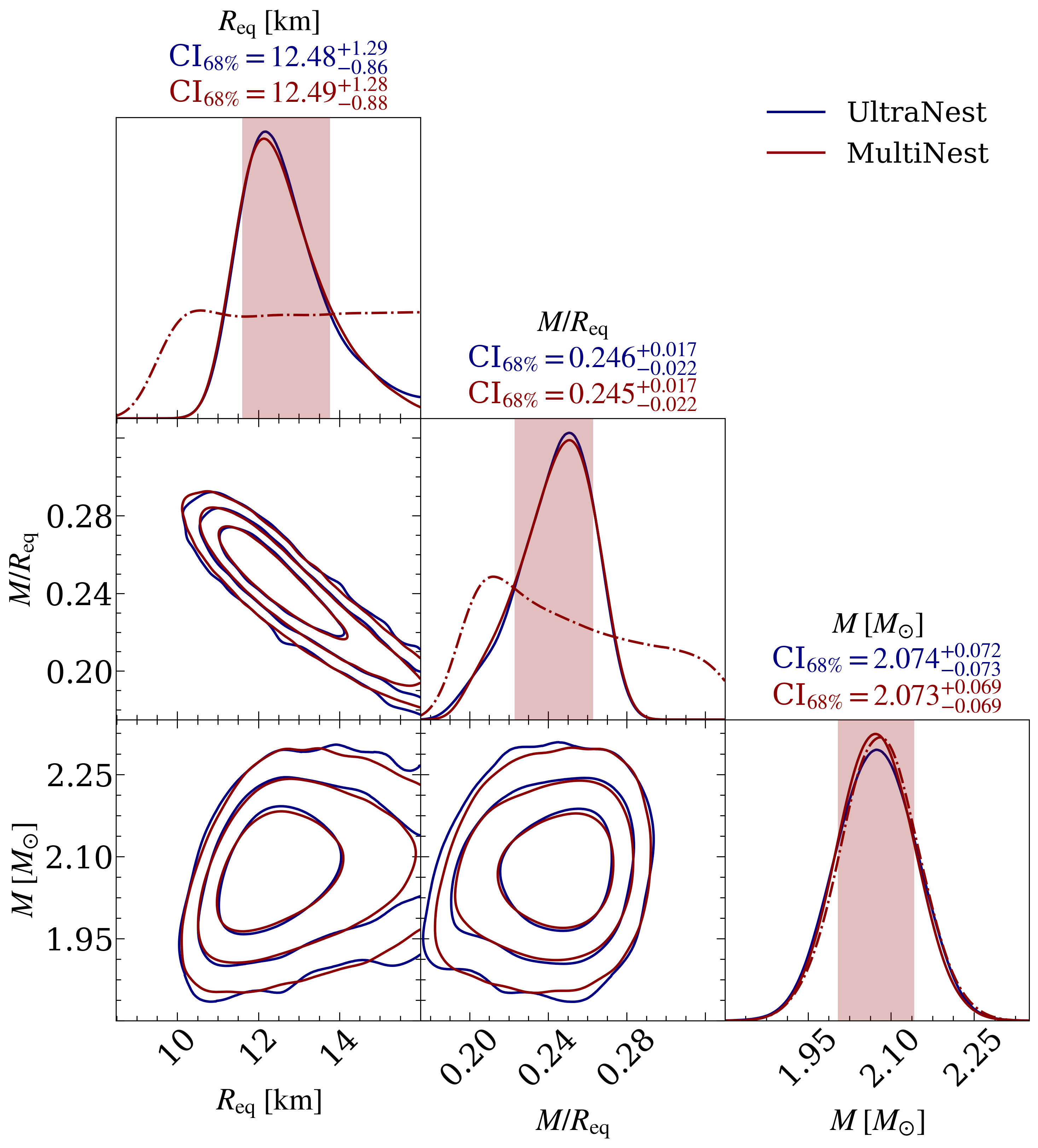}  
\caption{\small \justifying{
Radius, compactness, and mass posterior distributions using the PSR J0740+6620 joint NICER and XMM-Newton data set conditional on the \texttt{ST-U} model. Two posterior distributions are shown: the results from \cite{Salmi2024-J0740} using \texttt{MultiNest} using $4\times10^4$ live points and a sampling efficiency of 0.01, and the results obtained with the \texttt{UltraNest}'s slice sampling algorithm using a minimum of 1000 live points and 600 steps. The marginal prior PDFs for each parameter are displayed as dashed-dotted lines. The shaded regions in the diagonal panels contain the 68.3\% credible interval for each parameter symmetric around the median. The contours in the off-diagonal panels contain the 68.3\%, 95.4\%, and 99.7\% credible regions.}}
\label{fig:corner_small}
\end{figure}

\subsection{Implications for PSR J0740+6620} \label{subsec:inference} 
To investigate the implications of sampler choice on the inferred properties of PSR J0740+6620, we compare our inference results, obtained with two different sampling algorithms implemented in \verb|UltraNest|, to the headline findings of \cite{Salmi2024-J0740} using \verb|MultiNest|. The analysis of \cite{Salmi2024-J0740}, using \verb|MultiNest| with $4\times10^4$ live points and a sampling efficiency of 0.01, reported an equatorial radius of $12.49^{+1.28}_{-0.88}$ km and a gravitational mass of $2.073^{+0.069}_{-0.069}$ \Msun.

In our study, we conduct an inference run using \verb|UltraNest|'s slice sampler, employing a minimum of 1000 live points and 600 steps. To allow for finer sampling of the posterior distributions, the sampler settings for both \verb|MultiNest| and \verb|UltraNest| are more stringent than the exploratory settings used in the parameter recovery tests (see Section \ref{subsec:param_recovery}). 

The resulting posteriors from the joint NICER and XMM-Newton data (see Section \ref{sec:data}) are presented in Figure \ref{fig:corner_small} and \ref{fig:corner}. Notably, the posteriors match very well for both samplers, with overlapping credible intervals for all parameters. Only slight deviations are visible in the posterior tails (as seen in Figure \ref{fig:corner}), but these differences do not seem to significantly affect the inferred properties of PSR J0740+6620.

In terms of computational cost, the inference run with \verb|MultiNest| required 84k core hours, while \verb|UltraNest| required 319k core hours. Note these numbers are not an exact comparison because we utilized high-resolution X-PSI settings for \verb|MultiNest|, which are computationally more expensive \citep[see e.g.][]{Vinciguerra2023-J0030-sim, Choudhury2024-raytracing} as opposed to the low-resolution X-PSI settings used for \verb|UltraNest| (see Footnote \ref{footnote:res_settings}).

\begin{figure*}[ht!]
\centering
\includegraphics[
width=\textwidth]
{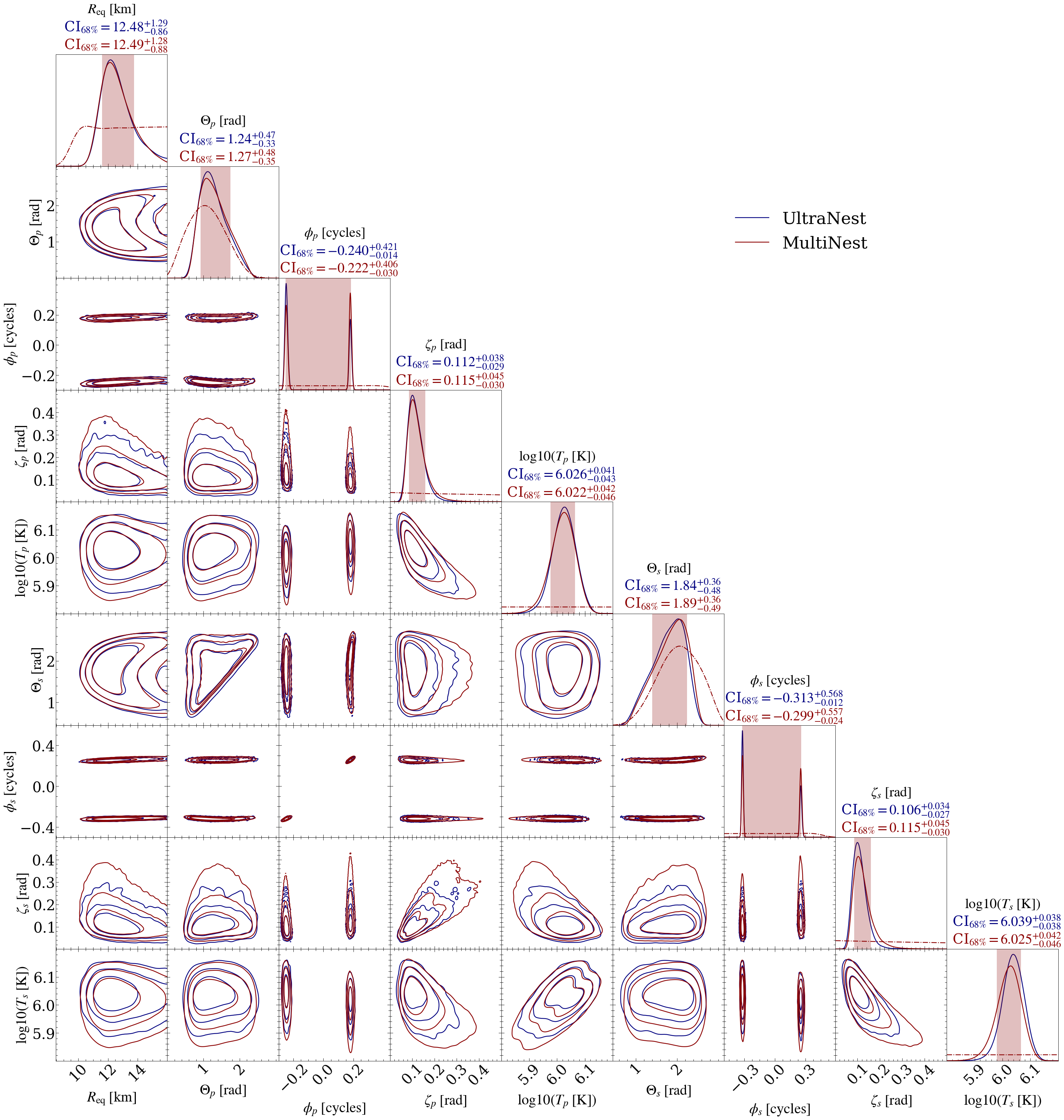}  
\caption{\small \justifying{
Posterior distributions for the hot region parameters using the PSR J0740+6620 joint NICER and XMM-Newton data set conditional on the \texttt{ST-U} model. See Figure \ref{fig:corner_small} for more details about the figure elements. Notably, the contours appear more ``wobbly" for \texttt{UltraNest} compared to \texttt{MultiNest}, which can be attributed to the smaller number of samples.}}
\label{fig:corner}
\end{figure*}

\section{Discussion \& Future Work} \label{sec:Discussion}
In this study, we used the X-PSI framework to assess the robustness of mass and radius inferences concerning our choice of sampler for the high-mass pulsar PSR J0740+6620, building on the work of \cite{Salmi2024-J0740}. We focused on two sampling algorithms: the multimodal nested sampling algorithm implemented in \verb|MultiNest| (used in \cite{Salmi2024-J0740}) and the slice nested sampling algorithm implemented in \verb|UltraNest|. Both samplers performed reliably, producing accurate and unbiased parameter estimation results when analyzing simulated data. Additionally, both samplers provided consistent results for the joint NICER and XMM-Newton PSR J0740+6620 data, with only slight deviations in the posterior tails (see Figure \ref{fig:corner}). This consistency shows that the choice of sampler does not significantly affect the inferred properties of PSR J0740+6620.

While our results demonstrated robustness in parameter recovery and inference, convergence remains an important consideration. For the real PSR J0740+6620 \verb|MultiNest| run, formal convergence has not been proven based on an assessment of the stability of the posteriors and evidence across multiple runs. Convergence is particularly challenging to establish for this source due to the flat likelihood surface at radii above $\sim 11$ km, which makes the upper limit difficult to constrain. Additionally, as the radius increases, the parameter space contains more solutions with smaller, hotter hot spots located closer to the poles (see Figure \ref{fig:corner}). In this region, the likelihood surface in spot size and temperature space exhibits a sharp, peaked structure. After additional exploration of this parameter space, \cite{Salmi2024-J0740} concluded that higher sampler settings — and consequently greater computational resources — are unlikely to significantly broaden the posterior because the volume is small and the overall maximum likelihoods in this restricted region remain lower than those in the broader prior space.

For the real PSR J0740+6620 \verb|UltraNest| run, convergence can be assessed in a similar manner as for \verb|MultiNest|. Prior test runs with lower sampler settings produced consistent results with Figures \ref{fig:corner_small} and \ref{fig:corner}, with the posterior and evidence remaining stable for an increasingly larger number of steps, indicating that the runs have converged. While the Mann-Whitney-Wilcoxon U-test (a feature in \verb|UltraNest|; see \cite{Buchner2023-NestedMethods, Fowlie2020-Utest}) detected significant deviations from uniform insertion orders, our series of runs varying the number of live points and steps yielded consistent posterior distributions and evidences, indicating that our results remain robust (see \cite{Fowlie2020-Utest} for similar situations).

It is important to note that the \verb|MultiNest| run on the real PSR J0740+6620 data utilized high-resolution X-PSI settings, whereas \verb|UltraNest| used low-resolution settings to reduce computational expense (see Footnote \ref{footnote:res_settings}). While high-resolution settings may be necessary in some cases for more accurate model calculations, prior tests suggest that this is unlikely to significantly affect the posterior \cite{Riley2021-J0740, Vinciguerra2023-J0030-sim, Choudhury2024-raytracing}. Consequently, the low-resolution settings used with \verb|UltraNest| are assumed to have a negligible impact on the final results.

\begin{table*}[btp]
\centering
\resizebox{\textwidth}{!}{ 
\begin{tabular}{@{}llllllllllll@{}} 
\hline\hline
\textbf{Parameter}                          & \textbf{Description}                                 & \multicolumn{10}{c}{\textbf{Injected values}} \\ 
\hline
$M$ $[\textit{M}_{\odot}]$                  & gravitational mass                                   & 2.126 & 2.092 & 2.113 & 2.166 & 2.014 & 2.198 & 2.054 & 2.025 & 2.038 & 2.109 \\
$R_{\textrm{eq}}$ $[$km$]$                  & coordinate equatorial radius                         & 12.176 & 11.019 & 10.197 & 11.528 & 12.141 & 9.839 & 12.040 & 10.742 & 10.421 & 9.730 \\
$D$ $[$kpc$]$                               & Earth distance                                       & 1.456 & 1.321 & 1.104 & 1.091 & 1.103 & 1.136 & 1.068 & 0.924 & 1.181 & 0.955 \\
$\cos(i)$                                   & cosine Earth inclination to spin axis                & 0.043 & 0.044 & 0.043 & 0.047 & 0.043 & 0.047 & 0.040 & 0.040 & 0.041 & 0.044 \\
$\phi_{p}$ $[$cycles$]$                     & $p$ region initial phase                             & 0.447 & -0.458 & 0.497 & -0.104 & 0.134 & 0.122 & -0.411 & 0.228 & 0.072 & -0.376 \\
$\Theta_{p}$ $[$radians$]$                  & $p$  region center colatitude                        & 0.862 & 0.723 & 0.711 & 1.795 & 1.897 & 0.663 & 0.490 & 1.476 & 0.478 & 1.399 \\
$\zeta_{p}$ $[$radians$]$                   & $p$ region angular radius                            & 0.661 & 0.485 & 0.542 & 0.098 & 0.294 & 0.651 & 0.682 & 0.758 & 1.206 & 0.598 \\
$\log_{10}\left(T_{p}\;[\textrm{K}]\right)$ & $p$ region \texttt{NSX} effective temperature        & 6.435 & 5.123 & 5.501 & 6.097 & 6.502 & 6.467 & 5.914 & 5.765 & 5.318 & 5.642 \\
$\phi_{s}$ $[$cycles$]$                     & $s$ region initial phase                             & -0.478 & 0.428 & -0.338 & 0.176 & 0.363 & 0.196 & -0.263 & -0.459 & 0.250 & -0.280\\
$\Theta_{s}$ $[$radians$]$                  & $s$  region center colatitude                        & 1.501 & 2.029 & 1.359 & 2.156 & 2.165 & 2.826 & 2.509 & 1.522 & 2.946 & 1.599\\
$\zeta_{s}$ $[$radians$]$                   & $s$ region angular radius                            & 0.516 & 0.604 & 1.133 & 0.067 & 0.055 & 1.298 & 0.536 & 0.261 & 1.000 & 0.970 \\
$\log_{10}\left(T_{s}\;[\textrm{K}]\right)$ & $s$ region \texttt{NSX} effective temperature        & 6.122 & 6.527 & 6.221 & 5.520 & 6.490 & 5.331 & 5.298 & 5.753 & 6.687 & 6.235 \\
$N_{\textrm{H}}$ $[10^{20}$cm$^{-2}]$       & interstellar neutral H column density                & 7.589 & 3.001 & 5.386 & 6.944 & 9.462 & 3.994 & 4.273 & 3.996 & 0.733 & 3.081 \\
Noise seed                                  & random seed number for Poisson fluctuations          & 566 & 475 & 733 & 420 & 680 & 463 & 885 & 104 & 554 & 156 \\ 
\hline
\end{tabular}}
\caption{\small \justifying{Table with injected values for each parameter, rounded to three decimal places, corresponding to data sets 1 to 10 (left to right). Parameters for the primary hot region are denoted with a subscript $p$ and the parameters for the secondary hot region with a subscript $s$. Additional details of the parameter descriptions and the prior PDFs can be found in Table I in \cite{Salmi2024-J0740}.}}
\label{table:1}
\end{table*}

The number of parameter recovery tests in this study was limited by computational constraints. Using a larger number of simulations may uncover biases not detected in this study. We note that \verb|MultiNest| seems to perform well, recovering the injected parameter values within statistically expected credible intervals, with $4\times10^3$ live points and a sampling efficiency of 0.1. However, this does not necessarily imply that these sampler settings are sufficient for converged results. Higher sampler settings are needed for the real PSR J0740+6620 data to fully explore the parameter space, specifically the pointy end of the posteriors in the radius-colatitude and spot size-temperature planes. Additionally, $4\times10^3$ live points and a sampling efficiency of 0.1 were found to be insufficient to produce converged results for simulated data in \cite{Salmi2024-J0740}. This suggests that the sampler settings used in the parameter recovery tests for \verb|MultiNest| potentially underestimate the credible intervals. Expanding these tests to include a broader range of simulated datasets would be necessary to provide a more comprehensive evaluation of sampler performance under diverse conditions.

In summary, we assessed the impact of sampler choice on pulse profile modeling inference using X-PSI, focusing on the specific case of PSR J0740+6620. This is a challenging inference problem, but it is important to test samplers since accurately establishing the width of the credible intervals is critical to attempts to constrain the dense matter equation of state. Our results show that while \verb|UltraNest| incurs significantly higher computational costs than \verb|MultiNest|, it demonstrates convergence with stable inference results for PSR J0740+6620, provided that low-resolution X-PSI settings generalize to high-resolution cases. In contrast, \verb|MultiNest| does not have formally proven convergence. Despite these differences, both samplers produce reliable parameter estimates, as confirmed by parameter recovery tests. Overall, the consistency of the credible intervals obtained with both samplers for the joint NICER and XMM-Newton PSR J0740+6620 data reinforces the robustness of the results from \cite{Salmi2024-J0740}.

In future work, we aim to extend this analysis to other pulsars, such as PSR J0030+0451 and PSR J0437-4715, to further evaluate the generalizability and robustness of our pulse profile modeling procedure using X-PSI.

\begin{acknowledgments}
M.H. and A.L.W. acknowledge support from the NWO grant ENW-XL OCENW.XL21.XL21.038 \textit{Probing the phase diagram of Quantum Chromodynamics} (PI: Watts). 
Y.K., T.S., and A.L.W. acknowledge support from ERC Consolidator grant No.~865768 AEONS (PI: Watts).  
The NWO Domain Science subsidized the use of the national computer facilities in this research.
Part of the work was also carried out on the HELIOS cluster including dedicated nodes funded via the above-mentioned ERC Consolidator grant.
This research has made use of data products and software provided by the High Energy Astrophysics Science Archive Research Center (HEASARC), which is a service of the Astrophysics Science Division at NASA/GSFC and the High Energy Astrophysics Division of the Smithsonian Astrophysical Observatory.

\textit{Software:} Cython \citep{Behnel2011}, GetDist \citep{Lewis2019}, GNU Scientific Library \citep{Galassi2009}, HEASoft \citep{heasoft2014}, Matplotlib \citep{Hunter2007}, MPI for Python \citep{Dalcin2008}, \verb|MultiNest| \citep{Feroz2009-Multinest}, nestcheck \citep{Higson2018-nestcheck-joss}, NumPy \citep{Walt2011}, \verb|PyMultiNest| \citep{Buchner2016-PyMultiNest}, Python/C language \citep{Oliphant2007}, SciPy \citep{Jones2001-scipy}, \verb|UltraNest| \citep{Buchner2021-UltraNest}, and X-PSI \citep{xpsi}.

\textit{Facilities:} NICER, XMM-Newton\\
\end{acknowledgments}

\appendix*
\section{Synthetic Data Parameters}\label{appendix_syndat}
We generate ten synthetic datasets for NICER and XMM-Newton using ten distinct parameter vectors, which are detailed in Table \ref{table:1}. These parameter vectors are obtained by uniformly sampling from the prior distributions employed in the joint NICER and XMM-Newton analysis of PSR J0740+6620 (see Table I of \cite{Salmi2024-J0740} for details about the priors). Note that the effective-area scaling factors, $\alpha_{\rm{NICER}}$ and $\alpha_{\rm{XMM}}$, are fixed to 1.0. The last column of Table \ref{table:1} shows the random seed used to generate Poisson fluctuations.

\bibliography{main}

\end{document}